\documentclass[aps,pra,twocolumn,epsfig,superscriptaddress,floats]{revtex4}
\usepackage{amsmath}
\usepackage{color}
\usepackage{physics}
\usepackage{graphicx}
\usepackage{amsfonts}
\usepackage{mathrsfs}
\usepackage{amsmath}
\usepackage{amssymb}
\usepackage{float}
\usepackage[utf8x]{inputenc}
\usepackage{times}
\usepackage{hyperref}
\usepackage{multirow}
\usepackage[normalem]{ulem}
\hypersetup{
  colorlinks=true,
  citecolor=blue,
  linkcolor=blue,
  urlcolor=blue}

\begin{document} 

\title{Classical route to ergodicity and scarring phenomena in a two-component Bose-Josephson junction}
 
\author{Debabrata Mondal}
\affiliation{Indian Institute of Science Education and
Research-Kolkata, Mohanpur, Nadia-741246, India}
\author{Sudip Sinha}
\affiliation{Indian Institute of Science Education and
Research-Kolkata, Mohanpur, Nadia-741246, India}
\author{Sayak Ray}
\affiliation{Physikalisches Institut, Rheinische Friedrich-Wilhelms-Universit\"at Bonn, Nussallee 12, 53115 Bonn, Germany}
\author{Johann Kroha}
\affiliation{Physikalisches Institut, Rheinische Friedrich-Wilhelms-Universit\"at Bonn, Nussallee 12, 53115 Bonn, Germany}
\author{Subhasis Sinha}
\affiliation{Indian Institute of Science Education and
Research-Kolkata, Mohanpur, Nadia-741246, India}
\date{\today}

\begin{abstract}

We consider a Bose-Josephson junction (BJJ) formed by a binary mixture of ultracold atoms to investigate the manifestation of coherent collective dynamics on {\it ergodicity and quantum scars}, unfolding the connection between them. By tuning the inter- and intra-species interaction, we demonstrate a rich variety of Josephson dynamics and transitions between them, which plays a crucial role in controlling the overall ergodic behavior. The signature of underlying classicality is revealed from the entanglement spectrum, which also elucidates the formation of quantum scars of unstable steady states and of periodic orbits leading to athermal behavior. The degree of ergodicity across the energy band and scarring phenomena can be probed from the auto-correlation function as well from the phase fluctuation of the condensates, which has relevance in cold atom experiments.

\end{abstract}


\maketitle

\section{Introduction} 

Coherent collective dynamics in a quantum many body system is a fascinating phenomenon. It attracts considerable interest since the realization of Bose-Josephson junctions (BJJ) formed by coupling two atomic Bose-Einstein condensates (BEC) in a double well trap \cite{Ketterle,Levy, Schmiedmayer, Oberthaler1, Oberthaler2, Oberthaler3,Oberthaler_review}. The interplay between phase coherence of the BECs and inter-particle interaction can lead to various non-linear oscillations \cite{Walls,Leggett1,Smerzi1,Smerzi2,Angela_Foerster,Kroha1,Kroha2}, quantum transitions \cite{Lewenstein, Oberthaler2, Oberthaler_review}, phase diffusion \cite{Vardi1, Leggett2, Oberthaler3, Stringari1} and onset of chaos \cite{Vardi_chaos_driven_BJJ,quantum_turbulence}.
Moreover, cold atom systems are an ideal platform to study such out-of-equilibrium phenomena of many-body systems \cite{Dalibard,Greiner_1,Greiner_2,Bloch_imaging, Greiner_3, quantum_gas_microscope_review, I_Bloch}. 
 
To date, a complete understanding of ergodicity and deviations from it in an interacting quantum systems remains a challenging issue. 
To shed light on it, the eigenstate thermalization hypothesis (ETH) \cite{Deutsch, Srednicki, P_Reimann} has been put forward, and its connections with the random matrix theory (RMT) has also been explored \cite{Izrailev, Santos_and_Rigol, Santos_and_Vyas, Santos_and_Izrailev, Polkovnikov_review}. 
However, deviation from ergodicity and the violation of ETH  has also drawn a lot of interest in the recent years \cite{Rigol_thermalization,MBL_review1}. In this context, a recent experiment on a chain of strongly interacting Rydberg atoms reveals that even in the ergodic regime, a special choice of initial states exhibits non-thermal and revival behavior \cite{Rydberg_chain_expt}. This has been attributed to many-body quantum scars (MBQS) \cite{Turner, scars_review, Abanin, Lin, Lukin1,Lukin2}, which has also been theoretically studied in other interacting models \cite{Schecter, AKLT_chains, Regnault1,Regnault2,Mark, scars_optical_lattice, correlated_bosons, Rainbow_scars, Onsager_scars, Ising_ladders, Fazio_ladder,Shane, scars_BJJ, coupled_top, kicked_coupled_top}.
The concept of quantum scar was originally introduced for single-particle states as the enhancement of spectral density near a quantum state whose corresponding classical orbit is unstable \cite{Heller}. However, such correspondence in a generic many-body system is not obvious due to the absence of a phase-space description. 
  
\begin{figure}
	\centering
	\includegraphics[width=\columnwidth]{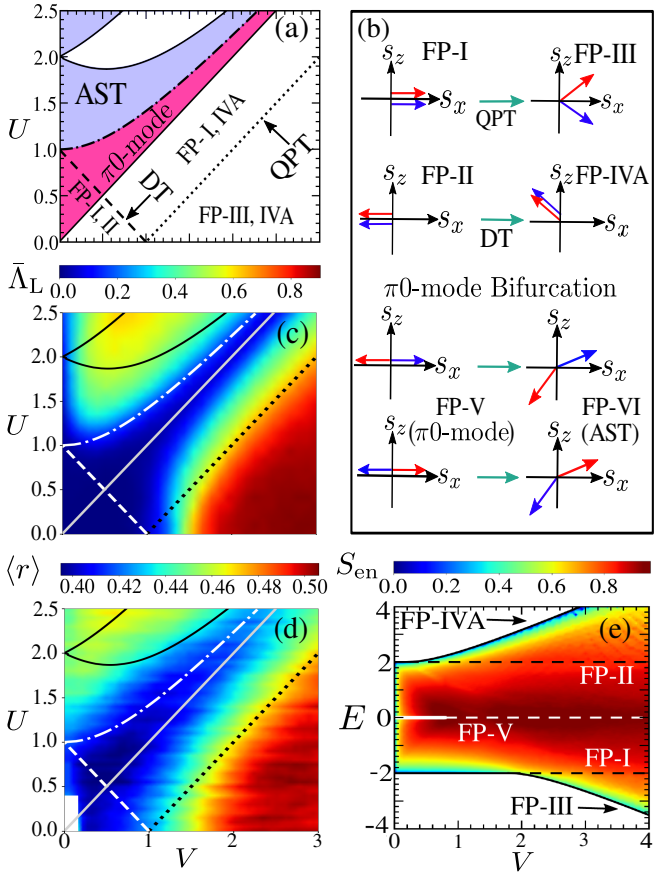}
	\caption{{\it Collective dynamics and ergodicity:}
	(a) Steady-state phase diagram as a function of interaction strengths $U$ and $V$ where the phase boundaries	are obtained from the stability analysis, see the text for details. The shaded regions denote the stability of the `$\pi 0$-mode' (dark magenta) and the AST state (light blue). The AST state becomes unstable in the white region (triangular area). (b) Schematics of spin orientation corresponding to different steady states (FP). Degree of ergodicity, quantified from the average (c) Lyapunov exponent $\bar{\Lambda}_{\rm L}$ and (d) ratio of level spacing $\langle r \rangle$, is shown as a color scale in the $U$-$V$ plane \cite{level_spacing_footnote}. The lines carry the same meaning as in (a). (e) Entanglement entropy $S_{\rm en}$ of the eigenstates scaled by the page value $S_{\rm max}$ \cite{Page} at energy density $E$,  with increasing $V$ at $U=0.8$. The solid (dashed) lines denote stable (unstable) FPs mentioned in the figure. For (d,e), $S=40$. Here and in all other figures, the inter(intra)-species interaction strength $V(U)$ and energy $E$ are measured in units of hopping amplitude $J$. We set $\hbar, k_{B}=1$.
	}      
\label{Fig1}
\end{figure}

This raises the question of a possible connection between underlying classicality and ergodicity of an interacting system, that can unveil an alternate route to ergodicity as well as formation of scars. To address this issue, we consider an experimentally realizable setup of a two-component BJJ \cite{2_comp_BJJ_1,2_comp_BJJ_2,2_comp_BJJ_3,2_comp_BJJ_4,
2_comp_BJJ_5}, where its phase coherence and collective nature can pave the way to explore such a connection. In this work, we demonstrate how a rich variety of steady states can influence the overall ergodicity of a BJJ, as summarized in Fig.\ref{Fig1}.  To investigate the route to ergodicity and its deviation, we explore the entanglement properties of such bipartite system in details. Apart from entanglement entropy, the  entanglement spectrum (ES) can also unveil the various features of a many body system starting from its topological aspects to many body localization \cite{ES_Topology1,ES_Topology2,ES_MBL}. In the present work, we demonstrate how the underlying classicality can also be unfolded from the ES, which elucidates the quantum scarring phenomena. 
Such feature of ES can also provide a deeper understanding of using the time dependent matrix product states with reduced dimensionality, which can shed light on formation of MBQS in a generic many body system \cite{Abanin,Lukin1}. Finally, we discuss the methods to probe the energy dependent ergodicity and scarring phenomena in this experimentally realizable setup.

The rest of the paper is organized as follows. We describe the model for two component BJJ in Sec.\ref{Section_model} and analyze the different branches of Josephson dynamics, their stability, as well as transitions between them in Sec.\ref{Section_classical}.  
In Sec.\ref{Section_Classical_Quantum}, the manifestation of underlying classicality in quantum ergodicity is discussed through spectral statistics and bipartite entanglement.
Next, we investigate the quantum scars of unstable fixed points and periodic orbits in Sec.\ref{Section_Quantum_scar}.  In Sec.\ref{Section_Dynamical_signature},  we discuss the methods to detect the scars as well as probe the energy dependent ergodicity by using auto-correlation function and phase diffusion. Finally, we summarize our results and conclude in Sec.\ref{Section_conclusion}.

\section{The model}
\label{Section_model}
The BJJ formed by a binary mixture of ultracold bosons with equal population $N$ of each component, can be described within two-mode approximation \cite{Walls} by the Hamiltonian,
\begin{equation}
\hat{\mathcal{H}}=\sum_{i,\alpha}\left[-\frac{J}{2}\hat{a}^\dagger_{i,\alpha}\hat{a}_{i,\bar{\alpha}}+\frac{U}{2N}\hat{n}_{i,\alpha}(\hat{n}_{i,\alpha}-1) 
+\frac{V}{2N}\hat{n}_{i,\alpha}\hat{n}_{\bar{i},\alpha}\right]
\end{equation}
The first two terms represent a two-site Bose-Hubbard model with on-site interaction strength $U$ and hopping amplitude $J$ between the two sites denoted by $\alpha\in \{{\rm L},{\rm R}\}$ ($\bar{\alpha}\neq \alpha$), where, $\hat{a}_{i,\alpha}(\hat{a}_{i,\alpha}^{\dagger})$ represents the annihilation (creation) operator of the two species of bosons indexed by $i\in \{1,2\}$ ($\bar{i}\neq i$). The third term describes the inter-species interaction of strength $V$. We set $\hbar, k_{B}=1$ and scale energy (time) by $J\ (1/J)$.

This Hamiltonian of a two-component BJJ can be written as a generalized coupled top model \cite{Robb, Ballentine} describing two interacting large spins [see appendix \ref{Appendix_derivation} for derivation],
\begin{eqnarray}
\hat{\mathcal{H}}=-\hat{S}_{1x}-\hat{S}_{2x}+\frac{U}{2S}\left(\hat{S}_{1z}^2+\hat{S}_{2z}^2\right)+\frac{V}{S}\hat{S}_{1z}\hat{S}_{2z}
\label{Hamiltonian}
\end{eqnarray}
where the spin components of each species with magnitude $S=N/2$ are written as $\hat{S}_{ix}=\sum_{\alpha} \hat{a}_{i\alpha}^{\dagger}\hat{a}_{i\bar{\alpha}}/2$ and $\hat{S}_z=(\hat{n}_{i{\rm L}}-\hat{n}_{i{\rm R}})/2$, within the Schwinger-Boson representation.

\section{Classical dynamics and steady states}
 \label{Section_classical}
For large spins, $S \!\! \gg \!\! 1$, the spin operators can be treated as the components of the classical spin vector, $\vec{S_{i}}=S(\sin{\theta_{i}}\cos{\phi_{i}},\sin{\theta_{i}}\sin{\phi_{i}},\cos{\theta_{i}})$. Consequently, the Hamiltonian in Eq.\ref{Hamiltonian} can be written in terms of the canonically conjugate variables $\phi$, $z=\cos{\theta}$ as, 
\begin{eqnarray}
\mathcal{H}_{\rm cl} = \sum_{i}\left(-\sqrt{1-z^2_{i}}\cos{\phi_{i}}+\frac{U}{2}z^2_{i}\right) + Vz_{1}z_{2}
\end{eqnarray}
Note that, the classical Hamiltonian $\mathcal{H}_{\rm cl}$ is scaled by $S$, and the corresponding classical energy $E$ is equivalent to the quantum mechanical energy density $\mathcal{E}_{n}/S$, where $\mathcal{E}_{n}$ are the energy eigenvalues of $\mathcal{\hat{H}}$ in Eq.\ref{Hamiltonian}. The corresponding classical equations of motion (EOM) are given by,
\begin{eqnarray}
\dot{z}_i = -\sqrt{1-z_i^2}\sin\phi_i,\quad \dot{\phi}_i = \frac{z_i\cos\phi_i}{\sqrt{1-z_i^2}}+Uz_i+Vz_{\bar{i}}\quad 
\label{EOM}
\end{eqnarray}	
Here, $z_i$ and $\phi_i$ denote the population imbalance $(n_{i{\rm L}}-n_{i{\rm R}})/N$ and the relative phase between the condensates on the two sites, for each atom-species $i$.
The resulting, rich variety of collective Josephson dynamics can be demonstrated on a Bloch sphere using the spin representation \cite{Oberthaler2,2_comp_BJJ_1}. We investigate them by analyzing the fixed points (FPs) of the above EOM.
A stability analysis performed around the FPs, $\{z^{*}_{1},z^{*}_{2},\phi^{*}_{1},\phi^{*}_{2}\}$, reveals transitions between the steady states as discussed below.

The symmetry unbroken ground state, FP-I: $\{z^{*}_{1,2}=0,\phi^{*}_{1,2}=0\}$ with energy density $E=-2$, becomes unstable when $V\geq U+1$ [dotted-line in Fig.\ref{Fig1}(a)], undergoes a quantum phase transition (QPT) and bifurcates to the two-fold degenerate, antiferromagnetic ground states FP-III: $\{z^{*}_{1}=-z^{*}_{2}=\pm\sqrt{1-1/(U-V)^2},\phi^{*}_{1,2}=0\}$ with $E=1/(U-V)+U-V$. 
Similarly, the highest excited state, FP-II: $\{z^{*}_{1,2}=0,\phi^{*}_{1,2}=\pi\}$ with $E=2$, becomes unstable when $V\geq-U+1$ [dashed-line in Fig.\ref{Fig1}(a)] and undergoes a dynamical transition (DT), giving rise to symmetry-broken ferromagnetic states, FP-IVA: $\{z^{*}_{1,2}=\pm\sqrt{1-1/(U+V)^2},\phi^{*}_{1,2}=\pi\}$ with $E=1/(U+V)+U+V$, see Fig.\,\ref{Fig1}(b).
 In the context of BJJ, the ferromagnetic and antiferromagnetic orders indicate the same and opposite density imbalance with equal
magnitude, respectively, corresponding to the two atomic
species \cite{2_comp_BJJ_1}. FP-IVA represents a self-trapped state \cite{Smerzi1,Smerzi2,Oberthaler2} with equal imbalance of both the species.
%

Additionally, there are two important class of steady states which include 
FP-V: $\{z^{*}_{i}=z^{*}_{\bar{i}}=0,\phi^{*}_{i}=0,\phi^{*}_{\bar{i}}=\pi\}$  ($i\neq\bar{i}$) with $E=0$. We call these states as `$\pi 0$-mode', describing the inter-species phase difference $\pi$, which is represented by an angle between the two spins, see Fig.\ref{Fig1}(b). It remains stable in the region $V<U<\sqrt{V^2+1}$ (magenta region in Fig.\ref{Fig1}(a)) and becomes unstable across the line $U=\sqrt{V^2+1}$ (dash-dotted line), bifurcating to `asymmetric self trapped' (AST) states, denoted by FP-VI: $\{|z^{*}_{i}|<|z^{*}_{\bar{i}}|\neq0,\phi^{*}_{i}=0,\phi^{*}_{\bar{i}}=\pi\}$ with unequal imbalance $|z_{1}|\neq|z_{2}|$ and $E>0$. This state remains stable in the light blue colored region of the phase diagram shown in Fig.\ref{Fig1}(a), whereas it becomes unstable in the white region (triangular area) in between the stable regions. Note that, in this white region, apart from the unstable AST state, there also exists other stable steady states: FP-I, IVA as well as unstable steady states: FP-II, FP-V ($\pi0$-mode) at different energies.
The spin orientations of FPs are summarized in Fig.\ref{Fig1}(b).
Next we ask, how such rich steady-state structure and the underlying collective dynamics influence the overall ergodicity of the system and what is the signature of emergence of such classicality in quantum dynamics?

\section{From classical to quantum ergodicity}  
\label{Section_Classical_Quantum}
In this section, we demonstrate the manifestation of underlying classicality on the overall quantum ergodicity.
In order to quantify the degree of chaos, we compute the Lyapunov exponent (LE) \cite{Strogatz,Lichtenberg} and obtain its mean value $\bar{\Lambda}_{\rm L}$ averaged over an ensemble of phase space points. The overall chaotic behavior in $U$-$V$ plane is displayed by $\bar{\Lambda}_{\rm L}$ as a color scale plot in Fig.\ref{Fig1}(c).
\subsection{Spectral statistics}
\label{Subsection_spectral}
In the quantum domain, the signature of chaos is studied from spectral statistics of the Hamiltonian in Eq.\ref{Hamiltonian}. We sort the eigenvalues $\mathcal{E}_{n}$ belonging to a particular symmetry sector of $\hat{\mathcal{H}}$ [see Appendix \ref{Appendix_spectral}]. To probe the degree of chaos, we compute the average level spacing ratio \cite{level_spacing_1}, namely, 
\begin{eqnarray}
\langle r \rangle = \langle {\rm min}(\delta_{n},\delta_{n+1})/{\rm max}(\delta_{n},\delta_{n+1}) \rangle
\end{eqnarray}
where $\delta_{n} = \mathcal{E}_{n+1}-\mathcal{E}_{n}$. In terms of $\langle r \rangle$, chaoticity is portrayed in the $U$-$V$ plane [cf. Fig.\ref{Fig1}(d)].
In the classically regular regime, the level spacing distribution follows Poisson statistics with $\langle r \rangle \sim 0.386$ \cite{level_spacing_2}. With increasing degree of chaoticity, $\langle r \rangle$ increases and finally approaches to $\langle r \rangle \sim 0.529 $ in the completely chaotic regime \cite{level_spacing_2}, where the underlying distribution of $\delta_{n}$ approaches Wigner-Surmise \cite{Haake}.

Remarkably, the map of dynamical chaos based on $\bar{\Lambda}_{\rm L}$ retains its fingerprints at the quantum level obtained from $\langle r \rangle$, see Fig.\ref{Fig1}(c,d). As expected, BJJ exhibits regular dynamics for weak interactions, whereas with increasing $V$, a crossover to chaos occurs when $V>U+1$. Interestingly, the stability of $\pi 0$-mode has a dramatic impact on the overall ergodicity of the BJJ, as evident from comparatively lower values of $\overline{\Lambda}_{\rm L}$ and $\langle r \rangle$ [cf. Fig.\ref{Fig1}(c,d)].  With increasing $U$, a mixed phase space behavior is observed above the region of stability of $\pi 0$-mode [see Fig.\ref{Fig1}(c)], where small regular islands form within the chaotic sea.

\subsection{Energy dependent ergodicity and mixed phase space}
\label{Section_Mixed_phase_space}
We also investigate the ergodic behavior of different eigenstates $\ket{\psi_{n}}$ across the energy band from relative entanglement entropy (EE) $S_{\rm en}/S_{\rm max}$, where 
\begin{eqnarray}
S_{en} = -{\rm Tr}(\hat{\rho}_{\mathcal{S}}{\rm ln}\hat{\rho}_{\mathcal{S}})
\end{eqnarray}
is computed from the reduced density matrix $\hat{\rho}_{\mathcal{S}} = {\rm Tr}_{\bar{\mathcal{S}}}\ket{\psi_{n}}\bra{\psi_{n}}$ obtained  by tracing out the other spin $(\bar{\mathcal{S}} \neq \mathcal{S})$. The degree of ergodicity is maximum at the center of the energy band with $E \approx 0$ compared to the band edges, indicating an energy dependent ergodic behavior [see Fig.\ref{Fig1}(e)]. Such behavior is also observed near delocalization to localization transition across the many body mobility edges \cite{mobility_edge1,mobility_edge2}. The maximum value of EE corresponding to a completely random state is given by \cite{Page},
\begin{eqnarray}
S_{\rm max} = {\rm ln}(2S+1)-1/2.
\end{eqnarray}
In the fully chaotic regime,  EE approaches to its maximum limit ($S_{\rm en} \simeq S_{\rm max})$ at the the band center. 

\begin{figure} 
	\centering
	\includegraphics[width=\columnwidth]{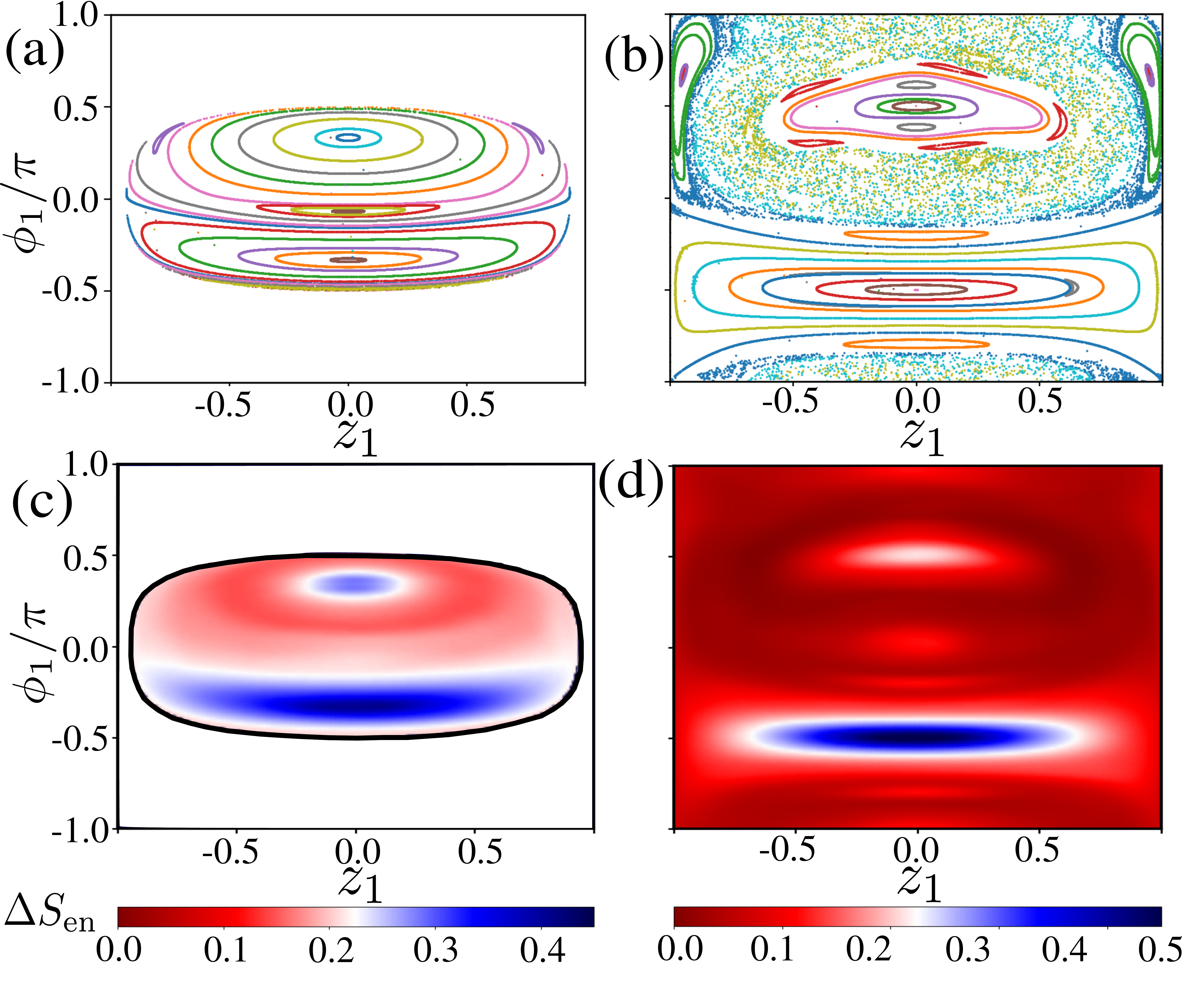}
	\caption{{\it Energy dependent degree of ergodicity:} Poincar\'{e} sections at $z_{2}=0$ plane for (a) $E = -1.0$ and (b) $E = 0.0$, for $U=0.8$ and $V=1.2$. (c,d) Color scaled plots of time averaged deviation of EE $\Delta S_{\rm en}$ from the ergodic limit for initial coherent states representing the same phase space points in (a,b) respectively. For quantum calculations, we set $S=30$.}
	\label{Fig2}
\end{figure}

To analyze such dynamical route to the ergodic behavior, we first plot the Poincar\'{e} sections at $z_{2}=0$ for different energies [cf. Fig.\ref{Fig2}(a,b)]. In quantum domain, we time evolve the initial coherent states $\ket{\psi_{c}}=\ket{z_{1},\phi_{1}} \otimes \ket{z_{2},\phi_{2}}$,
where $\ket{z,\phi}$ represents the spin coherent state given by \cite{coherent_state},
\begin{eqnarray}
\ket{z,\phi}=\left(\frac{1+z}{2}\right)^{S}\,\exp{\sqrt{\frac{1-z}{1+z}}e^{i\phi} \hat{S}_-}\ket{S,S}
\end{eqnarray}
which provides a semiclassical description of phase space points. Using the time-evolved state, we compute deviation of the late-time averaged EE from its maximum limit, namely, $\Delta S_{\rm en} = |\bar{S}_{\rm en}-S_{\rm max}|/S_{\rm max}$. As evident from Fig.\ref{Fig2}(c,d), the regular (chaotic) regions give higher (lower) $\Delta S_{\rm en}$, revealing the underlying classicality as well supporting the energy dependent ergodic behavior. 

\subsection{Underlying classicality from entanglement spectrum}
\label{Section_ES}
\begin{figure}
	\centering
	\includegraphics[width=\columnwidth]{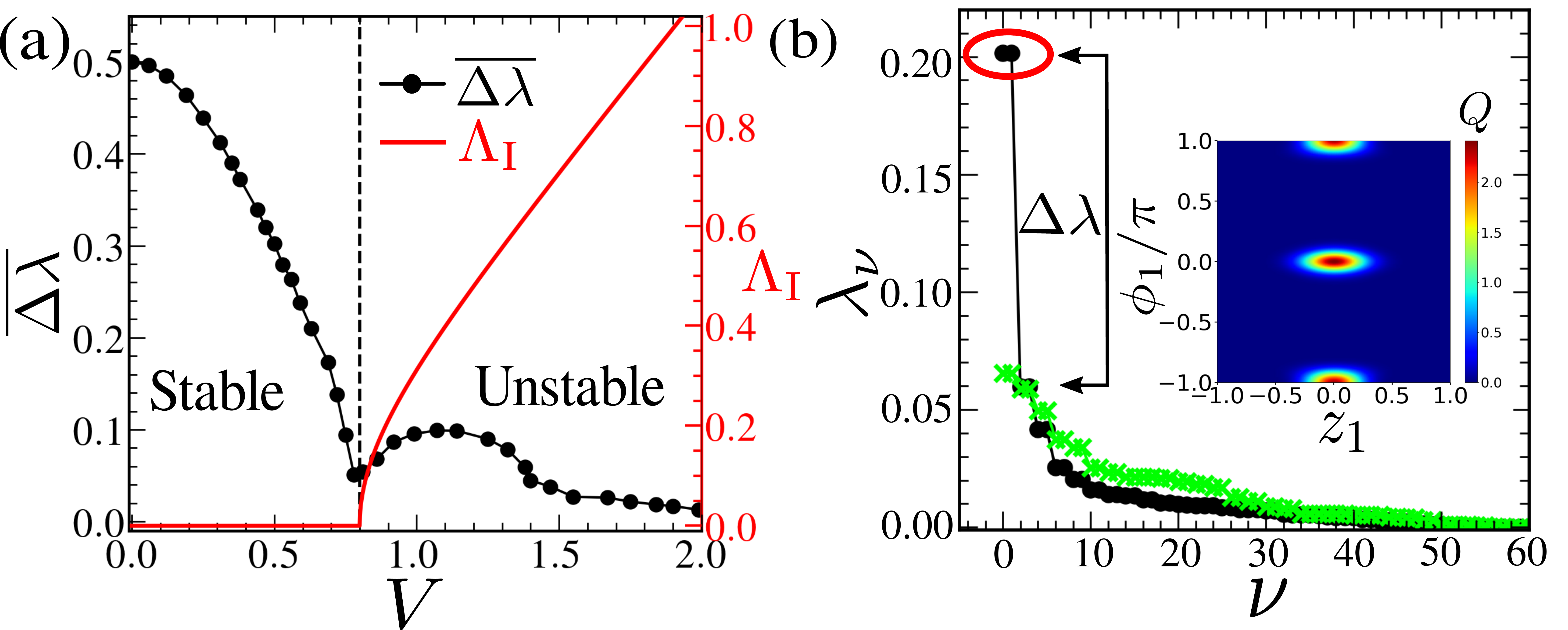}
	\caption{{\it Quantum scars of $\pi 0$-mode:} (a)Variation of time averaged eigenvalue gap $\overline{\Delta \lambda}$ (left axis) and instability exponent $\Lambda_{\rm I}$ scaled by $J$ (right axis) are shown for the $\pi 0$-mode with increasing $V$. (b) Entanglement spectrum (ES) for eigenstate containing the scar of $\pi 0$-mode exhibiting a gap $\Delta \lambda$ (black circles), and for arbitrary ergodic eigenstate (green crosses). In the inset, Husimi distribution $Q(z_{1},\phi_{1})$ of the reduced density matrix $\hat{\rho}^{\rm tr}_{S}$ corresponding to largest eigenvalues marked by red circle for the scarred eigenstate.}
	\label{Fig3}
\end{figure}
To study the signature of classicality in quantum dynamics, we  focus our discussion on the $\pi 0$-mode. We time evolve the initial coherent state $\ket{\pi_+} = \frac{1}{\sqrt{2}}(\ket{0,0} \otimes \ket{0,\pi} + \ket{0,\pi} \otimes \ket{0,0})$ describing the $\pi 0$-mode for sufficiently long time and study the `entanglement spectrum (ES)' of the final state $\ket{\psi(t)}$. The ES represents the eigenvalues $\{\lambda_{\nu} \}$ of the reduced density matrix,
\begin{eqnarray}
\hat{\rho}_S= \sum_{\nu}\lambda_{\nu} \ket{\nu}_S\bra{\nu}_{S}
\end{eqnarray}
which is obtained from the Schmidt decomposition of $\ket{\psi}=\sum_{\nu}\sqrt{\lambda_{\nu}}\ket{\nu}_{S}\otimes \ket{\nu}_{\bar{S}}$. 
In the weak interaction regime, for the stable $\pi 0$-mode, only a few  eigenvalues are significantly larger compared to others with a gap $\Delta\lambda$. 
Such structure of ES justifies the validity of product state in the weak coupling regime, capturing the classical dynamical behavior.
In contrast, the ES of an arbitrary ergodic state is extended and the eigenvalues $\{\lambda_{\nu}\}$ are distributed without any significant gap $\Delta \lambda$, which follows the `Marchenko–Pastur distribution' corresponding to Random matrix theory \cite{Pastur}. Such behavior can be observed in a strongly chaotic system \cite{Bandyopadhyay}. In the intermediate regime, we observe that the ES contains a few large eigenvalues separated from the extended tail by a gap $\Delta \lambda$ [see Fig.\ref{Fig3}(b)]. The distribution of eigenvalues in the tail part approaches to that of a random state. On the other hand, the reduced density matrix constructed from the few large eigenvalues of ES contains the underlying classical structure of phase space.
To investigate the dynamical signature of stability of the $\pi 0$-mode, we compute the time averaged gap $\overline{\Delta \lambda}$ for the final state $\ket{\psi(t)}$, with varying $V$ and compare it with the classical instability exponent $\Lambda_{\rm I}$ obtained from stability analysis [see also Appendix \ref{Appendix_steady_state}]. As seen from Fig.\ref{Fig3}(a), $\overline{\Delta\lambda}$ decays with increasing $V$ in the stable regime and a dip appears at the point of instability of the $\pi 0$-mode. 
Even after the instability of the $\pi 0$-mode, a few significantly large eigenvalues with a gap $\overline{\Delta \lambda}$ still persist in the ES [see Fig.\ref{Fig3}(a)], which retain the memory of the $\pi 0$-mode, leading to the formation of quantum scars. However as the system approaches the completely chaotic regime, the instability exponent of the $\pi 0$-mode grows rapidly. Consequently, the gap $\Delta \lambda$ vanishes as the distribution of eigenvalues $\{\lambda_{\nu}\}$ approaches to that of Marchenko–Pastur distribution.

\section{Quantum scars} 
\label{Section_Quantum_scar}
In this section, we investigate the quantum scarring phenomena which arises as a reminiscence of unstable fixed point and periodic orbit.
In the unstable regime of $\pi 0$-mode, we identify the scarred eigenstates $\ket{\psi_{n}}$ from a significant overlap with the coherent state $\ket{\pi_{+}}$ representing the $\pi 0$-mode, $|\langle \psi_n|\pi_+\rangle|^2 \gg 1/\mathcal{N}$, where $\mathcal{N}=(2S+1)^2$ is the system size. To illustrate the classicality of such scarred states, we construct a truncated reduced density matrix $\hat{\rho}^{\rm tr}_{S}$ corresponding to a few large eigenvalues in ES and compute the Husimi distribution,
\begin{eqnarray}
Q(z,\phi) = \frac{1}{\pi}\bra{z,\phi}\hat{\rho}^{\rm tr}_S\ket{z,\phi}
\end{eqnarray} 
which exhibits a localized phase-space density around FP-V ($\pi 0$-mode), indicating the scarring phenomena [cf. Fig.\ref{Fig3}(b) inset]. Additionally, there is another symmetry-broken antiferromagnetic state `FP-IVB', that exhibits scarring phenomena in the unstable regime, which is discussed in details in Appendix \ref{Appendix_scar_IVB}.

Apart from the fixed points, we also analyze the scars of periodic orbits in the most ergodic regime near $E \approx 0$. From EOM, we identify two classes of dynamics belonging to, Class-I:$\{z_1=-z_2, \phi_1=-\phi_2\}$ and Class-II:$\{z_1= z_2, \phi_1= \phi_2\}$, for which the dynamics is restricted in the respective sub-regions of the available phase space, containing two types of periodic orbits [see Appendix \ref{Appendix_periodic_orbits} for details]. 

\begin{figure}
	\centering
	\includegraphics[width=\columnwidth]{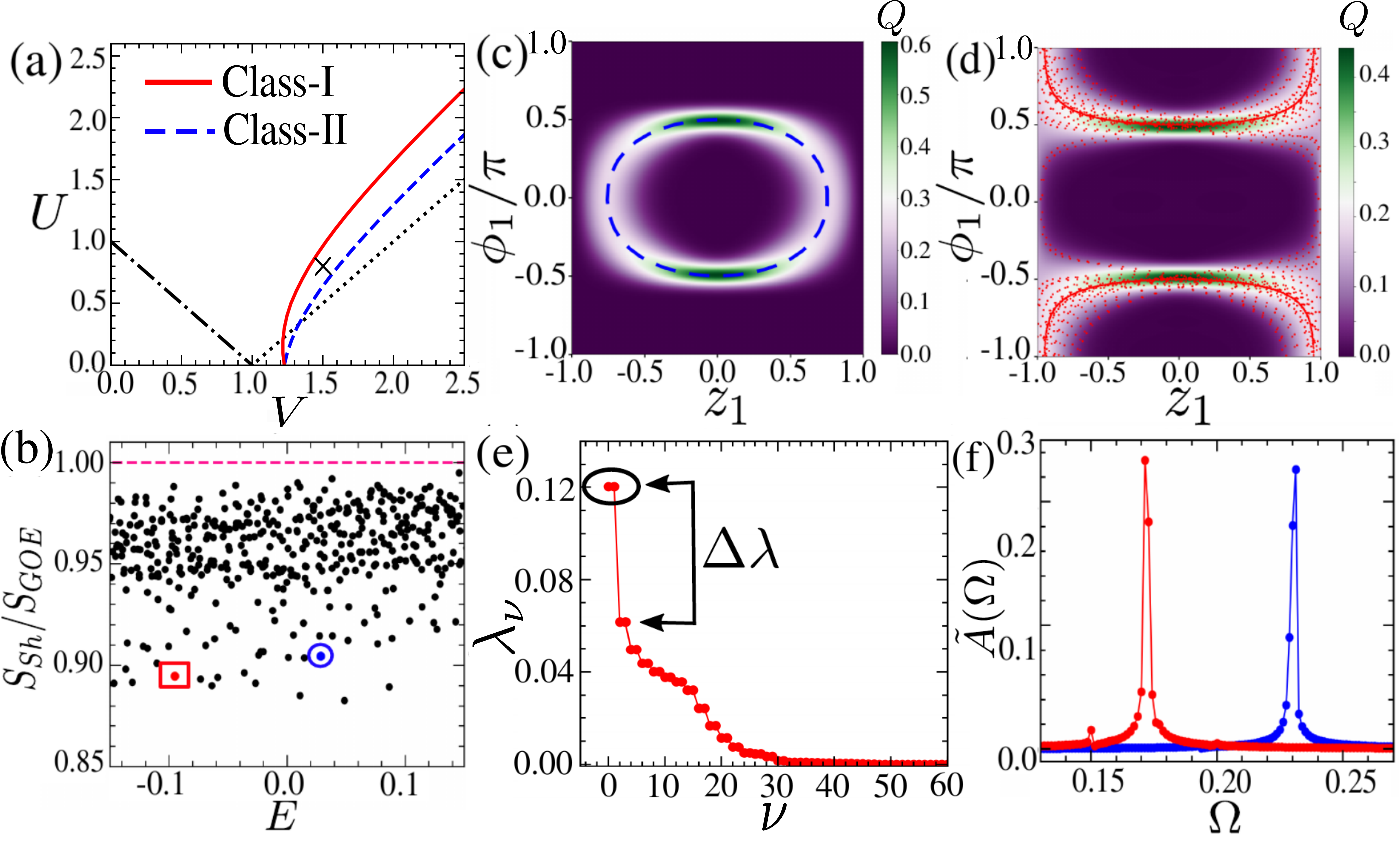}
	\caption{{\it Quantum scars of periodic orbits:}(a) Phase diagram of periodic orbits with $E=0$ belonging to different classes. The periodic orbits in class II (I) become unstable across the dashed blue (solid red) line with increasing $V$. (b) Shannon entropy of eigenstates $S_{\rm Sh}$ scaled by GOE value $S_{\rm GOE}$ near $E \approx 0$ at marked place in (a). (c-d) Husimi distribution $Q(z_{1},\phi_{1})$ of the encircled states in (b). The corresponding classical trajectories are ovelayed with initial conditions close to class II and I, respectively. (e) ES for eigenstates marked by red square in (b). (f) Fourier transform $\tilde{A}(\Omega)$ of auto-correlation function $A(t)$ evaluated for the encircled states in (b). Here and in remaining figures, time $t$ and frequency $\Omega$ are scaled by $1/J$ and $J$ respectively.}
	\label{Fig4}
\end{figure}
From stability analysis, we obtained a region in the parameter space, where the class-II orbits remain stable while the class-I orbits become unstable, see Fig.\ref{Fig4}(a). Correspondingly, we observe a few eigenstates in the ergodic regime ($E \approx 0$), maximally deviate from the GOE limit of the Shannon entropy, see Fig.\ref{Fig4}(b). Interestingly, amongst these deviated states, we identify the eigenstates bearing the scars of unstable orbits coexisting with those containing the image of the stable orbits, as evident from their respective Husimi distributions shown in Fig.\ref{Fig4}(c,d). 
Moreover, the scarred eigenstates exhibit a few significantly large eigenvalues separated from the rest by a gap $\Delta \lambda$ in the ES [see Fig.\ref{Fig4}(e)], retaining the memory of classical orbits in the phase space. To confirm the scarring due to periodic orbits, we compute the Fourier transform of the auto-correlation function \cite{Pollmann}, 
\begin{eqnarray}
A(t)=\sum_{a=x,y,z}\langle \hat{S}_{1a}(t)\hat{S}_{1a}(0) \rangle
\label{Auto_correlation}
\end{eqnarray}
evaluated for such scarred eigenstates. This exhibits a sharp peak at the frequency of the corresponding orbits [cf. Fig.\ref{Fig4}(f)], which can be obtained analytically [see Appendix \ref{Appendix_periodic_orbits}]. Note that, the scars of the periodic orbits have similar characteristics with that of the fixed points ($\pi0$-mode and FP-IVB). However, for the scar of a fixed point, the semiclassical phase space density (Husimi distribution) is localized around that point [see the inset of Fig.\ref{Fig3}(b)], whereas in the case  of periodic orbit, the density spreads out in phase space resembling the shape of the underlying classical orbit [see Fig.\ref{Fig4}(c,d)].

\section{Dynamical Detection of degree of ergodicity and scar}
\label{Section_Dynamical_signature}
\begin{figure}
	\centering
	\includegraphics[width=\columnwidth]{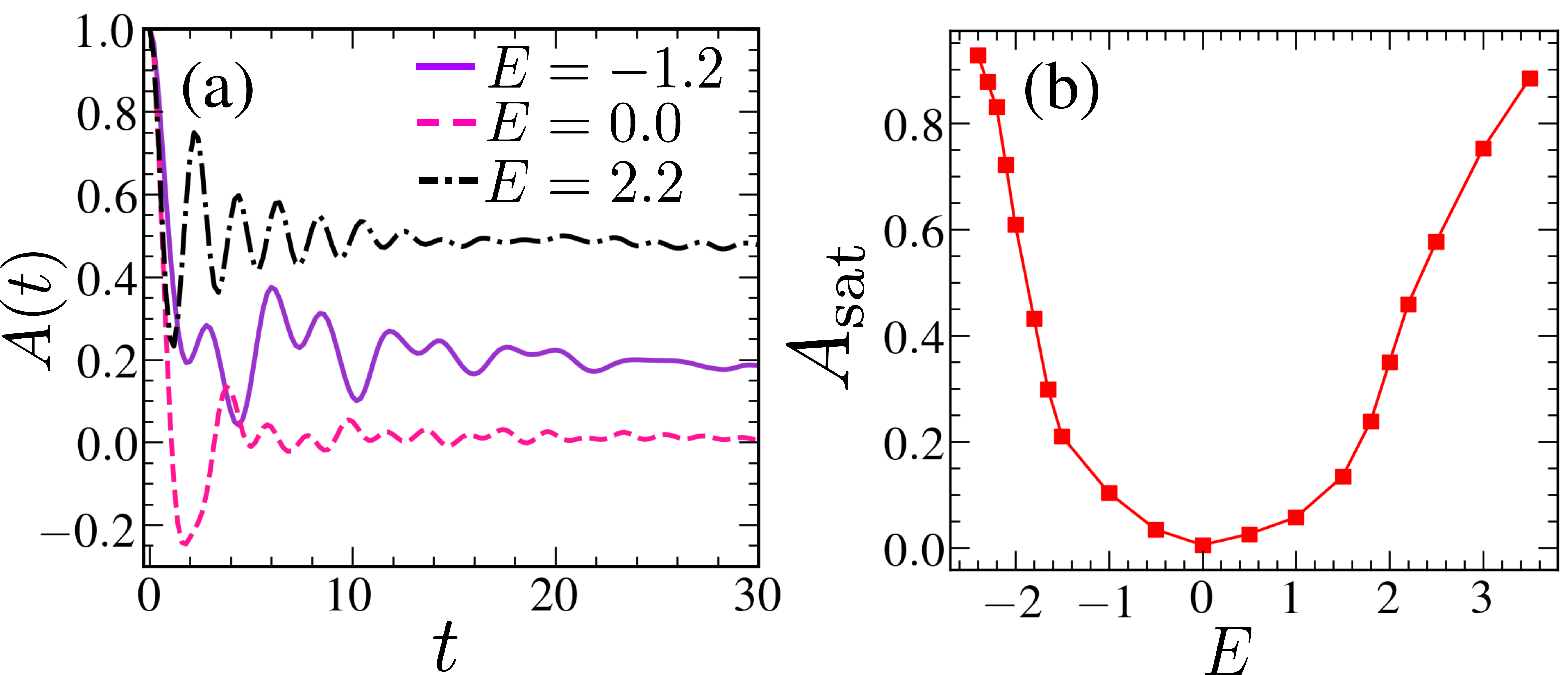}
	\caption{{\it Dynamical probing of energy dependent degree of ergodicity:} (a) Dynamics of auto correlation function $A(t)$ evaluated at different energy densities $E$ and for $U=0.8$, $V=2.8$. (b) Variation of saturation value of the auto correlation $A_{\rm sat}$ across the energy band.}
	\label{Fig5}
\end{figure}

To this end, we discuss the dynamical signature of energy dependent degree of ergodicity and quantum scar, from the auto-correlation function and phase diffusion dynamics.
The saturation value of auto-correlation function $A(t)$ [given in Eq.\eqref{Auto_correlation}], namely, $A_{\rm sat}$, averaged over an ensemble of initial coherent states $\ket{\psi_{c}}$ with fixed energy density $E$, can be used as a dynamical probe for the energy-dependent non-ergodic behavior. Notably, $A_{\rm sat}$ vanishes for states in the most ergodic region at $E\approx 0$, whereas it remains finite for the states near the band edges, see Fig.\ref{Fig5}.
The ergodic behavior can also be characterized by studying the phase coherence from non-equilibrium dynamics.
The phase coherence between the two sites of BJJ signifies the wave nature of the macroscopic condensate, and the relative phase between the two wells can be calculated by constructing an orthonormal basis of $2S+1$ phase states as following \cite{phase_operator, Oberthaler_review},
\begin{eqnarray}
\ket{\phi_m}=\frac{1}{\sqrt{2S+1}}\sum_{n=-S}^{S}\exp(in\phi_m)\ket{n}
\end{eqnarray}
with $\phi_m=\phi_0+2\pi m/(2S+1)$, where $m$ is an integer $m\in [0,2S]$ and  $\phi_m\in [-\pi,\pi]$. The phase distribution for a state $\ket{\psi}$ corresponding to a particular spin sector is given by $p(\phi_m) = {\rm Tr}(\hat{\rho}_{\mathcal{S}}\ket{\phi_{m}}\bra{\phi_{m}})$ with $\sum_{m}p(\phi_m)=1$, where $\hat{\rho}_{\mathcal{S}}={\rm Tr}_{\bar{S}}(\ket{\psi}\bra{\psi})$ is the reduced density matrix obtained by tracing out the other spin $(\bar{\mathcal{S}}\neq \mathcal{S})$. To study the phase diffusion dynamics, we evolve an initial coherent state with energy density $E$ and analyze its phase distribution corresponding to one of the bosonic component at different times. It is expected, the phase distribution for the states in the ergodic regime (center of the energy band with $E \approx 0$) will become flat quickly, indicating the loss of phase coherence. On the other hand, for the states in the non-ergodic regime (edge of the energy band), the phase distribution remains mostly localized and spreads comparatively less, indicating the retention of phase coherence.
\begin{figure}
	\centering
	\includegraphics[width=\columnwidth]{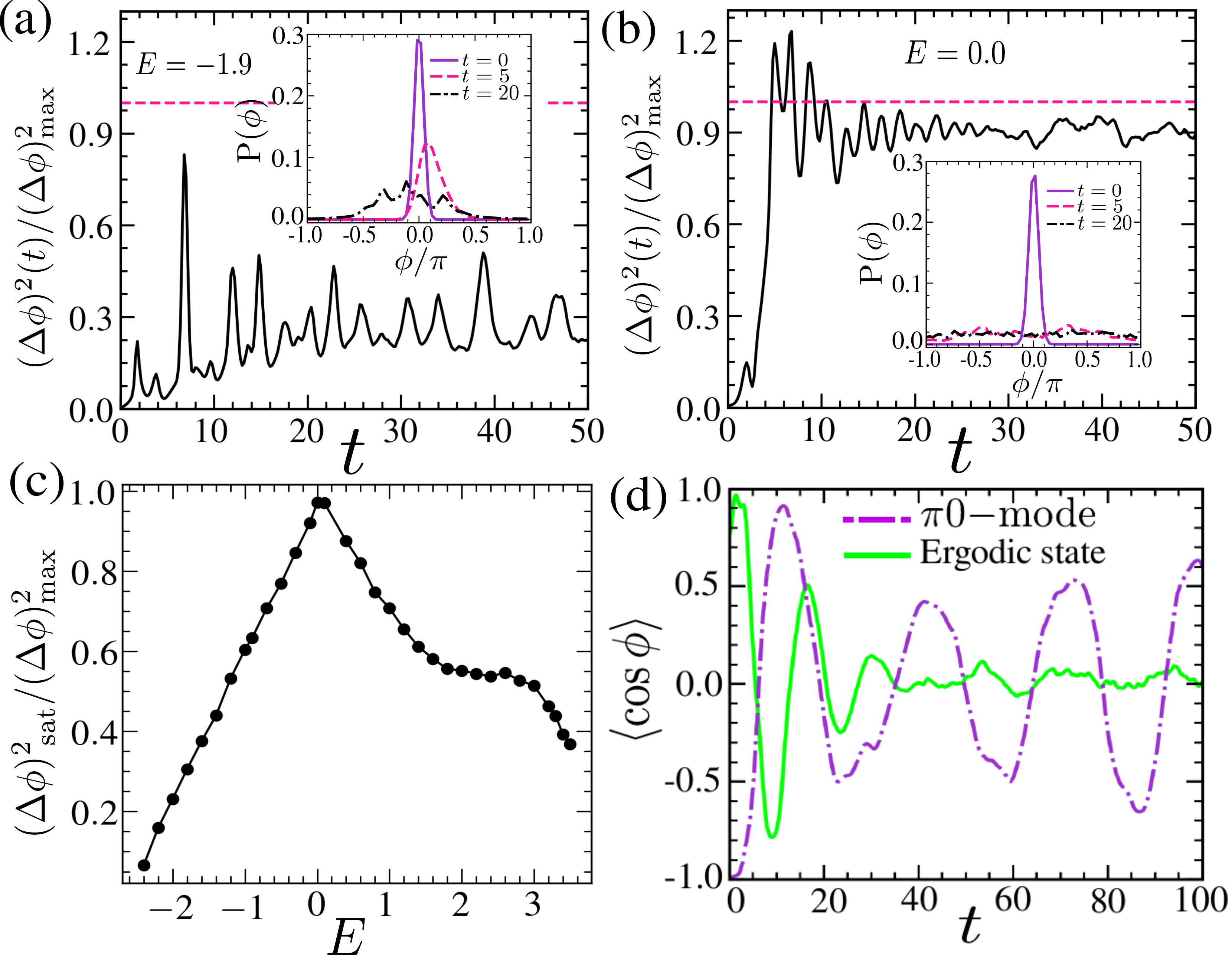}
	\caption{{\it Phase diffusion dynamics in two component BJJ}: Time evolution of the relative phase fluctuations $(\Delta \phi)^2(t)/(\Delta\phi)^2_{\rm max}$ (where $(\Delta\phi)^2_{\rm max} \simeq \pi^2/3 $) starting from  an arbitrary initial coherent state with energy density (a) $E=-1.9$ (non-ergodic) and (b) $E=0.0$ (ergodic) for $V=2.8$, indicating the energy dependent ergodic behavior. The insets show the snapshots of the corresponding phase distribution at different times. The horizontal dashed line indicates the maximum value $(\Delta\phi)^2/(\Delta\phi)^2_{\rm max} \simeq 1$. (c) Variation of saturation value of relative phase fluctuations $(\Delta\phi)^2/(\Delta\phi)^2_{\rm max}$ across the energy band. (d) Dynamics of the phase coherence factor $\langle \cos{\phi} \rangle$ for the scarred state of $\pi 0$-mode (dashed line) at $V=1.2$, exhibiting revival phenomena in contrast to an arbitrary ergodic state (solid line), which decays to zero, indicating loss of phase coherence. Parameters chosen: $S = 30, U = 0.8$.}
	\label{Fig6}
\end{figure}

To quantify the degree of ergodicity, we study the dynamics of phase fluctuations, 
\begin{eqnarray}
(\Delta\phi)^2 = \sum_{m}(\phi_m-\langle\phi\rangle)^2p(\phi_m)
\end{eqnarray}
for initial states at different energy densities $E$, where $\langle\phi\rangle = \sum_{m}\phi_{m}\,p(\phi_{m})$. As shown in Fig.\ref{Fig6}(a,b), starting from an initial state close to the band edge, the growth rate of $(\Delta\phi)^2$ is small and increases as we approach the band center ($E \approx 0$), which is already indicated from EE in Fig.\ref{Fig1}(e). In this regime, the phase fluctuation grows rapidly and saturates close to its maximal value $(\Delta\phi)^2_{\rm max}\simeq \pi^2/3$ corresponding to a random state \cite{Oberthaler_review}, indicating complete loss of phase coherence. In addition,  such behavior can also be quantified from $\langle \cos\phi\rangle = \sum_{m}\cos\phi_{m}\,p(\phi_m)$, where a (non) zero value signifies the loss (persistence) of the phase coherence. Interestingly, the phase coherence factor $\langle \cos{\phi} \rangle$ exhibits a revival phenomena for the unstable $\pi 0$-mode during the time evolution of the corresponding coherent state $\ket{\pi_+}$, capturing the scarring behavior. In contrast, for any other arbitrary state with same energy, $\langle \cos{\phi}\rangle$  decays to zero, signalling the loss of phase coherence. The energy dependent ergodicity can be probed and detection of quantum scarring of $\pi 0$-mode can be tested using relevant experiments \cite{Oberthaler3}.

\section{Conclusion}
\label{Section_conclusion}
To summarize, we have explored a rich variety of collective dynamics in a two-component BJJ, which unveils the classical route to ergodicity and quantum scarring phenomena in an interacting system. The steady states, particularly the $\pi0$-mode has a dramatic influence on the overall ergodic behavior. Moreover, an energy dependent ergodicity is also observed, and its connection with the mixed phase space regions is also explored.
As a signature of classicality, we identify a few significantly large eigenvalues in the entanglement spectrum (ES), which even exist for scarred eigenstates, retaining the memory of the unstable dynamics.
We demonstrate how the hidden classicality of a quantum state in dynamical evolution can be unfolded from the gap in the ES, which also persists for scarred states, separating a few large eigenvalues from the extended tail.  
The bipartite nature of BJJ makes it suitable to probe the above connection in terms of entanglement \cite{Greiner_3,ES_expt_PRL,ES_protocol}, which further addresses the issue of investigating the athermal dynamics in a reduced Hilbert space of a generic many body system \cite{Abanin,Pollmann}.
We elucidate the formation of scars of the fixed points as well as periodic orbits, and identified a region where the image of both the stable and unstable orbits can be observed in the Husimi distribution. In addition, we have also discussed methods to experimentally detect the scars as well energy dependent degree of ergodicity from phase diffusion \cite{Oberthaler3} and dynamics of auto-correlation.

Present work not only sheds light on underlying classicality of MBQS leading to athermal behavior, but the model can also be realized in spin systems with application to information processing \cite{Chudnovsky,A_Rey_expt,Bogani,
quantum_info_review} and lattice gauge simulation \cite{lattice_gauge}.

\begin{acknowledgments}

SR and JK acknowledge financial support by the Deutsche Forschungsgemeinschaft (DFG) through CRC/TR185 (277625399) OSCAR and through the Cluster of Excellence ML4Q (390534769). SR also acknowledges a scholarship from the Alexander von Humboldt Foundation, Germany.
\end{acknowledgments}

\appendix
\begin{table*}
	\begin{center}
		\renewcommand*{\arraystretch}{2.0}
		\begin{tabular}{|c| c c c c c|}
			\hline
			\large{$\phi_1^*,\phi_2^*$} & &\quad & \large{$z_1^*,z_2^*$} & &
			\\
			\hline
			&&\underline{FP-I :} & \underline{FP-III (Symmetry broken antiferromagnetic):} && \\
			0, 0 & & $z_1^*=z_2^*=0,$ &\,\,
			$z_1^*=-z_2^*=\pm\sqrt{1-\frac{1}{(U-V)^2}}$ & &\\
			&&Exists for all $U,V$&Exists for $U<V-1$&&\\
			&&Stable for $U>V-1$&Stable for $U<V-1$&&\\
			[3pt]
			\hline
			&&\underline{FP-II ($\pi\pi$-mode):} & \underline{FP-IVA (Ferromagnetic SST):} &\underline{FP-IVB (Antiferromagnetic SST):}& \\
			$\pi,\,\pi$ & &$z_1^*=z_2^*=0,$ & $z_1^*=z_2^*=\pm\sqrt{1-\frac{1}{(U+V)^2}}$, & $z_1^*=-z_2^*=\pm\sqrt{1-\frac{1}{(U-V)^2}}$ & \\
			&&Exists for all $U,V$&Exists for $U>-V+1$&Exists for $U>V+1$&\\
			&&Stable for $U<-V+1$&Stable for $U>-V+1$&Stable for $(U-V)^3\le U+V$&\\
			[5pt]
			\hline
			& &\multirow{1}{*}{\underline{FP-V ($\pi0$-mode):}} & \multirow{1}{*}{\underline{FP-VI (Asymmetric self-trapped):} }& &\\
			$0,\,\pi$ & &\multirow{1}{*}{$z_1^*=z_2^*=0$,} & $|z_1^*|<|z_2^*|$  & & \\
			$\pi,\,0$ & &\multirow{1}{*}{$z_1^*=z_2^*=0$,} &$|z_1^*|>|z_2^*|$ & &\\
			&&Exists for all $U,V$&Exists for $U>\sqrt{V^2+1}$&&\\
			&&Stable for $V<U<\sqrt{V^2+1}$&\qquad \quad See Fig.1(a) of the main text for stability&&\\
			\hline
		\end{tabular}
		\label{Tab:table}
		\caption{Chart of the fixed points (FPs) obtained from Eq.\ref{FP} corresponding to different steady states. Here SST stands for `symmetric self-trapped'. The region of existence and the stability of the steady states are also mentioned.}
	\end{center}	
\end{table*}
\section{Derivation of generalized coupled top model}
\label{Appendix_derivation}
In order to derive the effective spin Hamiltonian in Eq.2 from Eq.1 in the main text, we define the spin operators with spin $S=N/2$ within the Schwinger-Boson representation as follows, 
\begin{equation}
\hat{S}_{ix}=(a_{iR}^{\dagger}a_{iL}+a_{iL}^{\dagger}a_{iR})/2, \quad \hat{S}_{iz}=(\hat{n}_{iL}-\hat{n}_{iR})/2
\label{S-B_map}
\end{equation}
In this representation, the on-site and the inter-species interaction terms can be respectively written as,
\begin{eqnarray}
&&\frac{U}{2N}\,\sum_{\alpha}\hat{n}_{i\alpha}(\hat{n}_{i\alpha}-1) = \frac{U}{4N}\,  (N^2+4\hat{S}_{iz}^2)\\
&&\frac{V}{N}(\hat{n}_{1L}\hat{n}_{2L}+\hat{n}_{1R}\hat{n}_{2R}) = \frac{V}{2N}(N^2+4\hat{S}_{1z}\hat{S}_{2z})
\end{eqnarray}
Since each species have the equal number of population $\hat{n}_{iL}+\hat{n}_{iR}=N=2S$. Neglecting the zero point energy term, we can write the Hamiltonian of the two-component BJJ as generalized version of the coupled top model,
\begin{eqnarray}
\hat{\mathcal{H}}=-\hat{S}_{1x}-\hat{S}_{2x}+\frac{U}{2S}\left(\hat{S}_{1z}^2+\hat{S}_{2z}^2\right)+\frac{V}{S}\hat{S}_{1z}\hat{S}_{2z}\nonumber
\label{generalized CT}
\end{eqnarray}

\section{Steady states and their stability analysis}
\label{Appendix_steady_state}

Let us denote the steady states by the fixed points (FPs), $\mathbf{X}^* = \{z_1^* , z_2^*, \phi_1^*, \phi_2^* \}$. They are obtained by setting $\dot{z}_{i},\dot{\phi}_{i}=0$ in the equations of motion (EOM) in Eq.\ref{generalized CT},
\begin{subequations}
	\begin{eqnarray}
	\dot{z}_i &=& -\sqrt{1-z_i^2}\sin\phi^*_i=0 \label{phi} \\
	\dot{\phi}_i &=& \frac{z^*_i\cos\phi^*_i}{\sqrt{1-z_i^{*2}}}+Uz^*_i+Vz^*_{\bar{i}}=0
	\label{z}
	\end{eqnarray}
	\label{FP}
\end{subequations}
where, different species are denoted by $i\in\{1,2\}$ with $\bar{i} \ne i$. 
Note that, in a single component BJJ one observes three types of steady states, namely, `0-mode', `$\pi$-mode' and self-trapped state \cite{Smerzi1,Smerzi2}. Due to the inter-species interaction in the two-component BJJ, a hybridization between such steady states occurs, which gives rise to various different kinds of Josephson oscillations \cite{2_comp_BJJ_1}. Here, we present their stability analysis in details, which is important for the scarring phenomena associated with these steady states, charted in Table~I.
%
To investigate the stability of the steady states, we consider fluctuation around the FPs, namely, $\mathbf{X}(t) = \mathbf{X}^* + \delta \mathbf{X}(t)$, where $\delta \mathbf{X}(t) = \delta \mathbf{X} e^{i\omega t}$. By putting this in the EOM followed by an expansion upto a linear order in $\delta \mathbf{X}$, we obtain the fluctuation equations, 
\begin{subequations}
	\begin{align}
	&i\omega \delta z_i = \frac{z^{*}_i \sin\phi^*_i}{\sqrt{1-z_i^{*2}}} ~\delta z_i - \sqrt{1-z_i^{*2}}\cos\phi^*_i ~\delta \phi_i \label{dphi} \\
	&i\omega \delta \phi_i = -\frac{z^*_i\sin\phi^*_i}{\sqrt{1-z_i^{*2}}}~\delta \phi_i
	\!+\! \left[\frac{\cos\phi_i^*}{(1-z^*_i)^{3/2}} + U\right]\!\delta z_i \!+\! V\delta z_{\bar{i}} \label{dz}
	\end{align}
	\label{EOM_fluc}
\end{subequations}
The above set of equations in Eq.\ref{EOM_fluc} can also be represented as, $(\mathcal{J}-i\omega \mathbb{I})\delta \mathbf{X}=0$, where $\mathcal{J}$ is the Jacobian matrix and $\mathbb{I}$ is the identity. By solving this characteristic equation, we obtain the $\omega$ as,
\begin{eqnarray}
\omega_{\pm}^2 =\frac{A_1+A_2}{2} \pm \sqrt{\left(\frac{A_1-A_2}{2}\right)^2+B}
\end{eqnarray}
where $A_i = \left(\frac{\cos^2\phi_i^*}{1-z_i^{*2}}+U\sqrt{1-z_i^{*2}}\cos\phi_i^*\right)$ and $B = V^2\sqrt{1-z_1^{*2}}\sqrt{1-z_2^{*2}}\cos\phi_1^*\cos\phi_2^*$. The stability of the FP is ensured if $\omega$ is real and it represents the frequency of small amplitude Josephson oscillation. Whereas, for unstable FPs, the instability exponent is given by $\Lambda_{\rm I}=\Im[\omega]$, leading to an exponential growth of the fluctuation $\delta \mathbf{X}(t)$ over time.
%


\section{Symmetry classifications and spectral statistics}
\label{Appendix_spectral}
\begin{figure}
	\centering
	\includegraphics[width=\columnwidth]{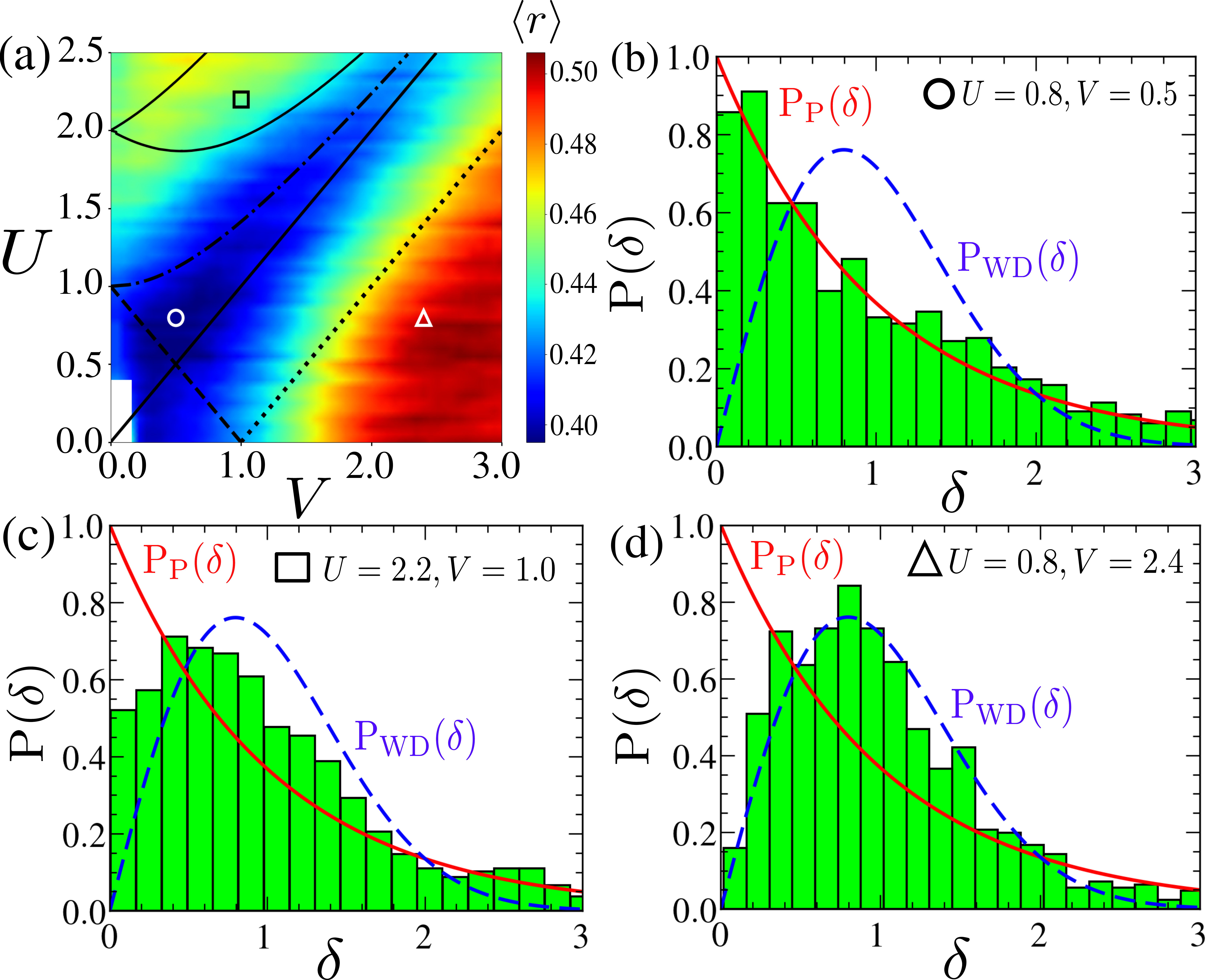}
	\caption{{\it Spectral analysis of energy levels}: (a) colormap of average level spacing $\langle r \rangle$ in the $U$-$V$ plane [see also Fig.1(d) in the main text]. (b-d) Level spacing distribution ${\rm P(\delta)}$ at the marked places are shown as histograms and are compared with the Poisson distribution ${\rm P_{\rm P}}(\delta)$ and with the Wigner-Surmise ${\rm P_{\rm WD}(\delta)}$, see the text for details. Here, we have set $S=40$ and $\delta$ is measured in units of $J$.}
	\label{fig:1_app}
\end{figure}

In order to study spectral statistics, we first compute the eigenspectrum of the effective spin Hamiltonian in Eq.\ref{Hamiltonian} by solving the following eigenvalue equation,
\begin{equation}
\hat{\mathcal{H}}\ket{\psi_{n}}=\mathcal{E}_n\ket{\psi_{n}}
\end{equation}
where, $\mathcal{E}_n$ are the eigenvalues and $\ket{\psi_{n}}$ are the associated eigenvectors. We note that the Hamiltonian $\hat{\mathcal{H}}$ has two symmetries---parity symmetry corresponding to the operator $\hat{\Pi} = e^{i\pi(\hat{S}_{1x} +\hat{S}_{2x} )}$ and spin exchange symmetry ($S_1 \leftrightarrow  S_2$) associated with operator $\hat{\mathcal{O}}$, that is constructed from $\bra{m_{1z},m_{2z}}\hat{\mathcal{O}}\ket{m_{2z},m_{1z}}=1$, where $m_{iz}$ are the quantum numbers of $\hat{S}_{iz}$. Both the operators, $\hat{\Pi}$ and $\hat{\mathcal{O}}$, have two eigenvalues, namely, $\pm$1. Accordingly, we separate out the eigenmodes into different symmetry sectors as follows,
\begin{eqnarray}
\langle \psi_{n}|\hat{\Pi}|\psi_{n}\rangle = \pm 1 \rightarrow \text{even(odd)}\nonumber\\
\langle \psi_{n}|\hat{\mathcal{O}}|\psi_{n}\rangle = \pm 1 \rightarrow \text{even(odd)}\nonumber
\end{eqnarray}
We focus on the eigenmodes belonging to the even-even symmetry sector. In Fig.\ref{fig:1_app}, we have plotted the distribution of the consecutive energy level spacings, $\delta_{n} = \mathcal{E}_{n+1}-\mathcal{E}_{n}$, with mean and normalization set to one \cite{Haake}, for various interaction strengths. Notably, in the stability region of the $\pi 0$-mode where the underlying dynamics is regular, the level spacing distribution ${\rm P}(\delta)$ exhibits Poisson statistics, ${\rm P_{P}}(\delta)=e^{-\delta}$ [see Fig.\ref{fig:1_app}(b)]. 
While above QPT, ${\rm P}(\delta)$ agrees with the Wigner surmise, ${\rm P_{WD}}(\delta)=\pi/2~\delta e^{-\pi\delta^2/4}$, corresponding to Gaussian Orthogonal Ensemble (GOE) [see Fig.\ref{Fig1}(d)] as a result of the onset of chaos in the phase space \cite{Haake}.
Interestingly, a mixed phase space is observed in the region of $U$-$V$ plane where the Lyapunov exponent acquires an intermediate value (particularly in the unstable region of`asymmetric self trapped (AST)' states, shown in Fig.1(a) of the main text),  which is reflected as an intermediate statistics of ${\rm P}(\delta)$, depicted in Fig.\ref{fig:1_app}(c).
\section{Classical periodic orbits}
\label{Appendix_periodic_orbits}
The spin exchange symmetry ($S_{1}\leftrightarrow S_{2}$) of the Hamiltonian in Eq.\eqref{Hamiltonian} gives rise to integrable motion restricted on a subregion of the phase space, where the dynamical variables satisfy the conditions: I. $\{z_{1}=-z_{2},\phi_{1}=-\phi_{2}\}$ and II. $\{z_{1}=z_{2},\phi_{1}=\phi_{2}\}$, defining the two dynamical classes. These conditions can equivalently be written as, I. $\{\phi_+=0;z_+=0\}$ and II. $\{\phi_-=0;z_-=0\}$, respectively, in terms of the new coordinates $z_\pm = (z_1 \pm z_2)/2$ and $\phi_\pm = (\phi_1 \pm \phi_2)/2$. Consequently, the dynamics of class I and II are governed by the EOM in terms of $\{z_{-},\phi_{-}\}$ and $\{z_{+},\phi_{+}\}$, respectively,
\begin{eqnarray}
\dot{z}_\pm&=&-\sqrt{1-z^2_\pm}\sin\phi_\pm\nonumber\\
\dot{\phi}_\pm&=&\frac{z_\pm}{\sqrt{1-z^2_\pm}}\cos\phi_\pm+(U\pm V) z_\pm
\label{equation_z-phi-pm}
\end{eqnarray}
The solution of above equations can be written in terms of elliptic functions as,
\begin{eqnarray}
z_\pm(t)&=&C_{\pm}\,\text{cn}\left(\frac{C_{\pm}\mu_{\pm}}{2k_{\pm}}(t+t_0),k_{\pm}\right)\nonumber\\
\cos(\phi_\pm(t))&=&-\frac{E+ \mu_{\pm} \,z^2_\pm(t)}{2\sqrt{1-{z^2_\pm(t)}}} \label{jacobi-elliptic-functions}
\end{eqnarray}
where, cn is the Jacobi elliptic function with elliptic modulus $k_{\pm}$ and the constants are defined in the following way [see also Ref.~\cite{Smerzi2}],
\begin{eqnarray}
C_{\pm}^2\,&=&\,\frac{2}{\mu_{\pm} ^2}\left[\frac{E\mu_{\pm} }{2}-1+\Omega_{\pm}\right], k_{\pm}^2=\frac{1}{2}\left[1+\frac{ E\mu_{\pm}/2 -1}{\Omega_{\pm}}\right]\nonumber\\
t_0 &=&\frac{F\left(\cos^{-1} (z_\pm(0)/C_{\pm}),k_{\pm}\right)}{\Omega_{\pm}^{1/2}}, \Omega_{\pm} =\sqrt{\mu_{\pm}^2+1- E\mu_{\pm}} \nonumber
\label{elliptic-modulus}
\end{eqnarray}
where, $\mu_{\pm} = (U\pm V)$ correspond to class-II and I respectively, and  $\text{F}(\phi,k_{\pm})\,=\,\int_{0}^{\phi}dx (1-k_{\pm}^2\sin^2 x)^{-1/2}$ is the incomplete elliptic integral of first kind.
For repulsive interactions, i.e. $U>0$ and $V>0$, the dynamics corresponding to the class-II describes an effective antiferromagnetic Lipkin-Meshkov-Glick (LMG) model \cite{LMG}. However, the dynamics of class-I can represent ferromagnetic or antiferromagnetic LMG model for $U-V<0$ and $U-V>0$, respectively.
Interestingly, even when the symmetry unbroken state FP-I (FP-II) becomes unstable in the full phase space, it can remain stable under the constraint of the corresponding dynamical class-I (class-II). 
Similarly, the periodic orbits forming around these fixed points with different energies $E$, can become unstable in presence of small fluctuation violating the conditions of the respective classes, depending on the strength of the interactions. The time period of the orbits with energy $E = 0$ belonging to the two classes is given by,
\begin{equation}
T = \frac{4K(k_{\pm})}{(1+\mu_{\pm}^2)^{1/4}} 
\label{time-period}
\end{equation}
where $K(k_{\pm})\,=\,\text{F}(\pi/2,k_{\pm})$.
The stability analysis of such periodic orbits is performed by using the method of Monodromy matrix described in \cite{Lichtenberg,monodromy2} and the stability regions of orbits with energy $E=0$ in the $U$-$V$ plane are shown in Fig.4(a) of the main text.

\section{Scar of Antiferromagnetic symmetric self-trapped state}
\label{Appendix_scar_IVB}

\begin{figure}
	\centering
	\includegraphics[width=\columnwidth]{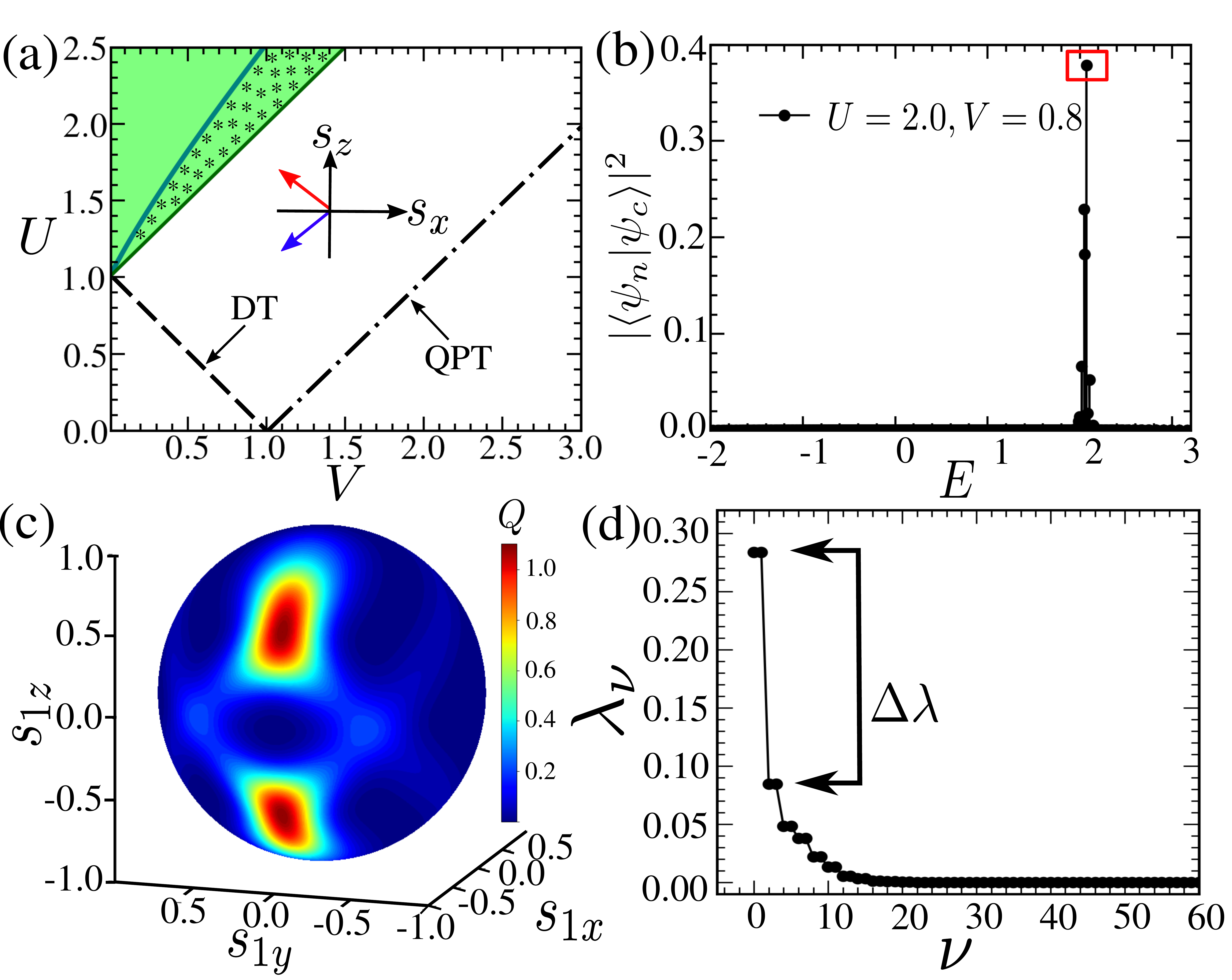}
	\caption{{\it Quantum scars of antiferromagnetic symmetric self-trapped state} (FP-IVB): (a) Phase diagram in the $U$-$V$ plane showing the presence of FP-IVB in the colored region along with the schematic of its spin configuration. The plain ($*$ shaded) region indicates the region of stability (instability). (b) Overlap $|\langle \psi_{n} | \psi_{c} \rangle|^2$ of the scarred eigenstates $\ket{\psi_{n}}$ with the coherent state $\ket{\psi_{c}}$ of the corresponding FP. (c) Husimi distribution  plotted on the Bloch sphere and (d) the entanglement spectrum for the eigenstate with maximum overlap marked by red square in (b).}
	\label{fig:2_app}
\end{figure}

As mentioned earlier, there exists a pair of symmetric self-trapped (SST) steady state FP-IVB with antiferromagnetic ordering in the regime $U\ge V+1$, with energy $E=\frac{1}{(U-V)}+U-V$ [see Fig.\ref{fig:2_app}(a)]. 
This state is originated after pitchfork bifurcation of symmetry unbroken state FP-II, when the dynamics is constrained within the class-I [as discussed previously in appendix \ref{Appendix_periodic_orbits}].
Even though this antiferromagnetic state is stable only in class-I, it looses stability in presence of small fluctuations violating the corresponding dynamical class, when $(U-V)^3\le U+V$, leading to the formation of scars. Such scarring phenomena can be analyzed quantum mechanically by the method described in the main text. We identify the corresponding scarred eigenstate $\ket{\psi_{n}}$ from the maximum overlap $|\langle \psi_n|\psi_c\rangle|^2 \gg 1/\mathcal{N}$ with the coherent state $\ket{\psi_{c}}$, describing FP-IVB semiclassically, as shown in Fig.\ref{fig:2_app}(b). As evident from Fig.\ref{fig:2_app}(c), the Husimi distribution of the scarred state shows localization of phase space density around the phase space point of the antiferromagnetic SST state, indicating the scarring phenomena. From the analysis of the entanglement spectrum (ES) of such scarred state [see the discussion in subsection \ref{Section_ES} of the main text], we also find a few large eigenvalues which are separated from the rest with a significant gap $\Delta \lambda$ [see Fig.\ref{fig:2_app}(d)], retaining classicality of the corresponding steady state. Such signature of quantum scar can also be probed experimentally.


\begin{thebibliography}{99}
\bibitem{level_spacing_footnote}
In the colorscale plot of $\langle r \rangle$, we kept the small region near the origin blank, since it is difficult to obtain a reliable statistics close to the non interacting regime ($U=0$, $V=0$).
\end{thebibliography}
\end{document}